\documentclass[conference]{IEEEtran}

\usepackage{url,hyperref,lineno,microtype,subcaption}
\usepackage{mathtools}
\usepackage{float}
\usepackage{graphicx}
\usepackage{subcaption}
\usepackage{booktabs,parskip}
\usepackage{algorithm}
\usepackage{algorithmic}
\usepackage{dsfont}
\usepackage[utf8]{inputenc}
\usepackage{dblfloatfix}
\usepackage{cite}
\usepackage{subcaption}
\usepackage{float}
\usepackage{url,hyperref,lineno,microtype,subcaption}
\usepackage{mathtools}
\usepackage{float}
\usepackage{graphicx}
\usepackage{subcaption}
\usepackage{booktabs,parskip}
\usepackage{algorithm}
\usepackage{algorithmic}
\usepackage{dsfont}
\usepackage[utf8]{inputenc}
\usepackage{graphics}
\usepackage{amsmath,amssymb,amsfonts}
\usepackage{algorithmic}
\usepackage{graphicx}
\usepackage{array}
\newcolumntype{P}[1]{>{\centering\arraybackslash}p{#1}}
\usepackage{textcomp}
\usepackage{xcolor}
\usepackage{ulem} 
\usepackage{textcomp}
\usepackage{mathtools}
\usepackage{authblk}
\def\BibTeX{{\rm B\kern-.05em{\sc i\kern-.025em b}\kern-.08em
    T\kern-.1667em\lower.7ex\hbox{E}\kern-.125emX}}

\usepackage[numbers]{natbib}
\usepackage{hyperref}
    
\begin{document}

\title{Beat-to-beat AV nodal assessment]{ECG-based beat-to-beat assessment of AV node conduction properties during AF (Preprint)}
{\footnotesize \textsuperscript{}}

}
\author[1,2]{Mattias Karlsson}
\author[2]{Felix Plappert}
\author[3]{Pyotr G Platonov}
\author[4]{Sten Östenson}
\author[1]{Mikael Wallman}
\author[2]{Frida Sandberg}

\affil[1]{Department of Systems and Data Analysis, Fraunhofer-Chalmers Centre, Sweden}
\affil[2]{Department of Biomedical Engineering, Lund University, Sweden}
\affil[3]{Department of Cardiology, Clinical Sciences, Lund University, Sweden}
\affil[4]{Department of Internal Medicine and Department of Clinical Physiology, Central Hospital Kristianstad, Sweden}

\maketitle

\begin{abstract}
The refractory period and conduction delay of the atrioventricular (AV) node play a crucial role in regulating the heart rate during atrial fibrillation (AF). Beat-to-beat variations in these properties are known to be induced by the autonomic nervous system (ANS) but have previously not been assessable during AF. Assessing these could provide novel information for improved diagnosis, prognosis, and treatment on an individual basis. \newline
\indent
To estimate AV nodal conduction properties with beat-to-beat resolution, we propose a methodology comprising a network model of the AV node, a particle filter, and a smoothing algorithm. The methodology was evaluated using simulated data and using synchronized electrogram (EGM) and ECG recordings from five patients in the intracardiac atrial fibrillation database. The methodology's ability to quantify ANS-induced changes in AV node conduction properties was evaluated by analyzing ECG data from 21 patients in AF undergoing a tilt test protocol. \newline
\indent
The estimated refractory period and conduction delay matched the simulated ground truth based on ECG recordings with a mean absolute error ($\pm$ std) of 169$\pm$14 ms for the refractory period in the fast pathway; 131$\pm$13 ms for the conduction delay in the fast pathway; 67$\pm$10 ms for the refractory period in the slow pathway; and 178$\pm$28 ms for the conduction delay in the slow pathway. These errors decreased when using simulated ground truth based on EGM recordings. Moreover, a decrease in conduction delay and refractory period in response to head-up tilt was seen during the tilt test protocol, as expected under sympathetic activation. \newline
\indent
These results suggest that beat-to-beat estimation of AV nodal conduction properties during AF from ECG is feasible, with different levels of uncertainty, and that the estimated properties agree with expected AV nodal modulation.

\end{abstract}


\section{Introduction} \label{Sec:intro}
\noindent
Atrial fibrillation (AF), characterized by disorganized electrical activity in the atria, is the most common sustained cardiac arrhythmia with an estimated prevalence between 2\% and 4\% globally \cite{benjamin2019heart}. The disorganized electrical activity in the atria leads to rapid and irregular contraction of the atria and ventricles, resulting in an increased risk of mortality, predominantly due to heart failure or stroke \cite{andrew2013prevalence}. Despite extensive research on AF, very little robust evidence exists to inform the best type and intensity of rate control treatment on an individual level \cite{ESC_Guidelines,al2014rate}.\newline
\indent
The atrioventricular (AV) node normally functions as the sole electrical connection between the atria and ventricles. During AF, the AV node plays a crucial role in protecting the ventricles from the rapid and irregular impulses originating in the atria. This function is accomplished through two distinct pathways; the fast pathway (FP) and the slow pathway (SP), which converge at the Bundle of His \cite{kurian2010anatomy}. The FP conducts impulses faster than SP but has a longer refractory period. Depending on the refractoriness of its pathways, the AV node can either block an incoming impulse or send it through with a conduction delay. Therefore, the refractory period and conduction delay of the two pathways -- here denoted ${\boldsymbol{\phi}} =$ [$R^{FP}$, $R^{SP}$, $D^{FP}$, and $D^{SP}$] -- are critical determinants of its filtering capability. The AV node thus serves an essential role in regulating the heart rate during AF, and can functionally be characterized by its properties ${\boldsymbol{\phi}}$.
\newline
\indent
The autonomic nervous system (ANS) has been shown to contribute to the initiation and maintenance of AF \cite{shen2014role}, suggesting that inter-patient variability in ANS activity might influence individual responses to AF treatment. During normal sinus rhythm, the ANS affects the heart rate primarily through changes to the sinus node automaticity, which can be quantified using heart rate variability \cite{shaffer2017overview}. However, during AF, the disorganized electrical activity in the atria overrides the organized electrical signals from the sinus node, preventing it from regulating the heart rate. Instead, the ANS affects the heart rate primarily through changes to the atrial fibrillatory rate and AV node conduction properties. Therefore, heart rate variability is not applicable as a tool for quantifying ANS modulation during AF. As an alternative, modulation of AV nodal function could be used to quantify the ANS function during AF. Since the AV node function mainly depends on the refractory period and conduction delay of the two pathways, estimating beat-to-beat changes to these properties might give insights into the ANS function. \newline
\indent
Assessing the AV-nodal function under AF is a complex task, since its behavior is influenced by multiple factors such as atrial impulses, autonomic modulation, as well as its intrinsic dynamics and structure. Thus, standard signal processing tools are insufficient, and a model-based analysis is required. Several mathematical models of the AV node have previously been proposed with varying complexity tailored to different intents, including but not limited to \cite{jorgensen2002mathematical,mangin2005effects,inada2009one,climent2011functional,mase2015dynamics,henriksson2015statistical,ryzhii2023compact}. For clinical application on an individual level, a model should ideally have parameters identifiable from non-invasive recordings. To the best of our knowledge, the only model incorporating the refractory period and conduction delay of both pathways while simultaneously allowing for identification of model parameters based on non-invasive recordings is our previously proposed model \cite{karlsson2021non}. Using this model, the individual 24-hour trends of the AV node properties ${\boldsymbol{\phi}}$ have previously been estimated with a temporal resolution of 10 minutes using an error function based on the Poincaré plot of the RR interval series \cite{karlsson2022ecg,karlsson2024model}. However, the ANS is known to modulate AV node conduction with beat-to-beat resolution \cite{leffler1994rate}. Thus, beat-to-beat resolution of the AV node conduction property estimates would be preferable for studying the ANS. \newline
\indent
Because the Poincaré plot relies on statistical information gathered over a sequence of multiple heartbeats, it is of limited use for beat-to-beat analysis. To increase temporal resolution, we propose a particle filter to estimate ${\boldsymbol{\phi}}$ with beat-to-beat resolution using our previously proposed model of the AV node \cite{karlsson2021non}. Particle filters approximate the solution to the filtering problem -- estimating the current state of a system (${\boldsymbol{\phi}}$ in our case) based on past and current observations. Due to their ability to leverage information from previous time points effectively, particle filters are suitable for beat-to-beat estimation. Particle filters have proven especially powerful for nonlinear and non-Gaussian problems, and have previously been used for e.g. atrial flutter detection \cite{lee2013atrial}, to robustly track heart rate \cite{nathan2017particle}, and to automatically annotate ultrasound videos of the fetal heart \cite{bridge2017automated}. Moreover, by combining the resulting estimates from a particle filter with a smoothing algorithm, estimates of the current state of a system based on past, current, and future observations can be obtained \cite{chopin2020introduction}. \newline
\indent This study aims to present and evaluate two particle filter-based frameworks for beat-to-beat assessment of AV node conduction properties, based on intracardiac electrogram (EGM) and ECG data, respectively. The evaluation is done in three steps. Step one is to evaluate the estimation accuracy for the EGM and ECG-based methods on simulated data. Step two is to compare the estimates obtained from synchronized ECG and EGM measurements to estimates derived from ECG measurements only. Finally, step three is to analyze the dynamics of the AV node properties during a tilt test protocol using ECG recordings from 21 patients in order to evaluate the method's ability to quantify expected changes in AV node conduction properties.

\section{Materials and Methods} \label{Sec:method}
The data used in this study are described in Section \ref{Sec:3_1}, and are followed by a description of the signal processing used to derive an atrial activation time series (AA series) and a ventricular activation time series (RR series) from the EGM and ECG recordings, in Section \ref{sec:rr_aa}. Furthermore, the network model of the AV node is described in Section \ref{Sec:3_3}, and the computation of the posterior distribution of $\boldsymbol{\phi}$ using a particle filter is described in Section \ref{Sec:PF_known} and \ref{Sec:PF_unknown}. Finally, the posterior estimates are refined using a smoothing algorithm, as described in Section \ref{Sec:smooth}.

\begin{figure*}[b]
    \centering
        \includegraphics[width=17cm]{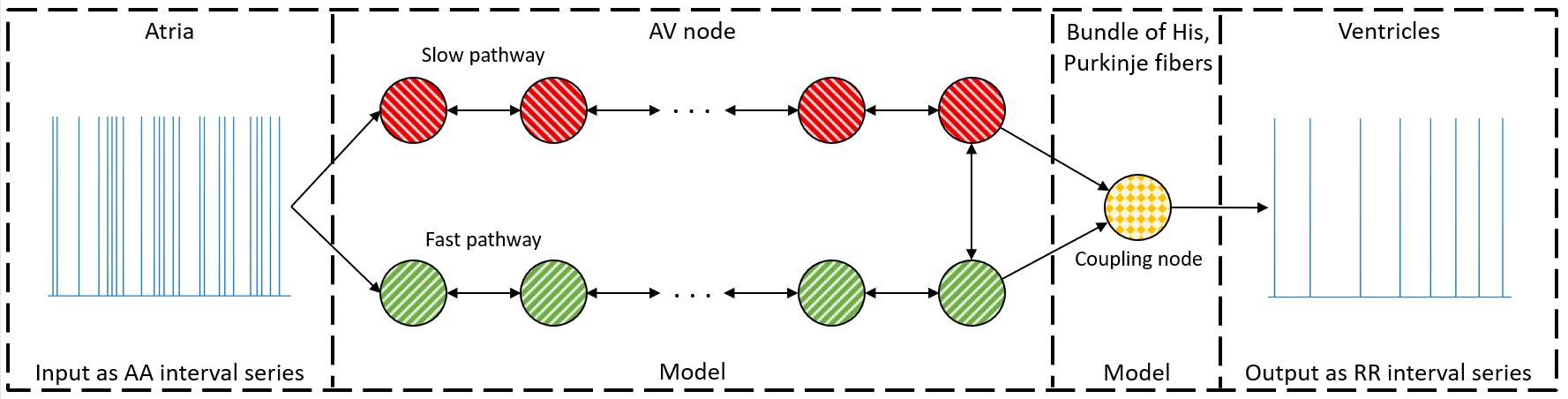}
    \caption{A schematic representation of the network model where the yellow node represents the coupling node, the red nodes the SP, the green nodes the FP, and arrows the direction for impulse conduction. For readability, only a subset of the 21 nodes is shown \cite{karlsson2021non}.}
    \label{Figure1}
\end{figure*}

\subsection{Datasets}\label{Sec:3_1}
\noindent
Two previously obtained datasets are used in this study. The publicly available intracardiac atrial fibrillation database (iafdb) provides synchronized EGM and ECG measurements during AF and is used to assess coherence between EGM and ECG-based estimates of $\boldsymbol{\phi}$ \cite{Goldberger2000}. Additionally, ECG recordings from a previously conducted tilt test study are used to study modulation of the AV node properties in response to changes in ANS activity \cite{ostenson2017autonomic}.
\newline
\subsubsection{Intracardiac atrial fibrillation database} \label{Sec:iafdb}
\noindent
The iafdb data consists of EGM recordings from four separate regions of the right atrium with synchronized three-lead ECG from eight patients with AF or flutter, sampled at 1000 Hz \cite{Goldberger2000}. The recordings at the tip of the tricuspid valve annulus are used in this study, due to its proximity to the AV-node entrance. Five recordings contain solely AF and were selected for analysis, with an average patient age of 73 $\pm$ 10 years, 60\% male, and an average signal duration of 58 $\pm$ 7 seconds. In addition, recordings with the catheter resting against the atrial free wall are used to create realistic simulated data, described in Supplementary Material S1. \newline

\subsubsection{Tilt test study} \label{Sec:tilt}
\noindent
The tilt test study includes ECG recordings from 40 patients with persistent AF \cite{ostenson2017autonomic}. For the current study, data with sufficient quality from 21 patients were used (average age of 67 $\pm$ 7 years, and 67\% male). Eleven patients were excluded due to missing ECG data, five due to the inability of the CardioLund software to detect R peaks (see Section \ref{sec:rr_aa_iafdb}) and three due to failure to extract f-waves (see Section \ref{Sec:AA:tilt}). The tilt test protocol involved standard 12-lead ECG recordings taken between 1 and 3 PM in a quiet room. Participants transitioned from supine position after approximately five minutes to head-down tilt (HDT) position (-30°) for approximately five minutes before finally a head-up tilt (HUT) position (+60°) for approximately five minutes. 

\subsection{Signal processing} \label{sec:rr_aa}
\noindent 
The frameworks presented in this study for assessing the AV node conduction properties with beat-to-beat resolution rely on simultaneous analysis of the RR and AA series. These are obtained using different signal processing methods depending on whether synchronized EGM and ECG recordings are available, or only ECG recordings, as described below.

\subsubsection{RR and AA series from synchronized EGM and ECG} \label{sec:rr_aa_iafdb}
\noindent 
The synchronized ECG and EGM recordings from the iafdb were used to derive the RR and AA series. The RR series is extracted using R-peak detection performed by the CardioLund ECG parser (\url{www.cardiolund.com}). The AA series is extracted from the EGM recordings using an iterative method \cite{ng2013iterative} following average beat subtraction-based ventricular far-field cancellation and standard pre-processing \cite{shkurovich1998detection, botteron1995technique}.

\subsubsection{RR and AA series from ECG} \label{Sec:AA:tilt}
\noindent Using solely ECG, the RR series is again extracted from the R-peak detection performed by the CardioLund ECG parser. However, the AA series cannot be extracted from the ECG. Instead, multiple AA series are generated for each RR interval based on the f-wave characteristics of the corresponding ECG segment. Each AA series, denoted $\boldsymbol{\alpha}$, is generated by a Gaussian random walk described by a mean ($\mu^{\alpha}$) and standard deviation ($\sigma^{\alpha}$). Both $\mu^{\alpha}$ and $\sigma^{\alpha}$ are estimated based on the f-wave signal extracted from the ECG by applying QRST-cancellation using the CardioLund ECG parser, before a harmonic model \cite{henriksson2018model} is fitted to the f-waves to estimate the f-wave frequency and a signal quality index ($SQI$), sampled at 50 Hz, as described in \cite{abdollahpur2022subspace}. For each RR interval, the mean ($\mu^{f}$) and standard deviation ($\sigma^{f}$) of the inverse f-wave frequency are calculated, as well as the $SQI$. Subsequently, $\mu^{\alpha}$ is drawn from ${\mathcal{N}}( \mu^{f}, \text{max}(0, 0.3-SQI)^4)$, where a $SQI$ greater than 0.3 is deemed sufficient based on previous studies \cite{henriksson2018model}, and the factor 4 is chosen to get a quadratic decrease on the variance for $SQI$ below 0.3. Further, $\sigma^{\alpha}$ is set to $4 \sigma^{f}$, where the factor 4 is chosen empirically.

\begin{figure*}[b]
    \centering
        \includegraphics[width=17cm]{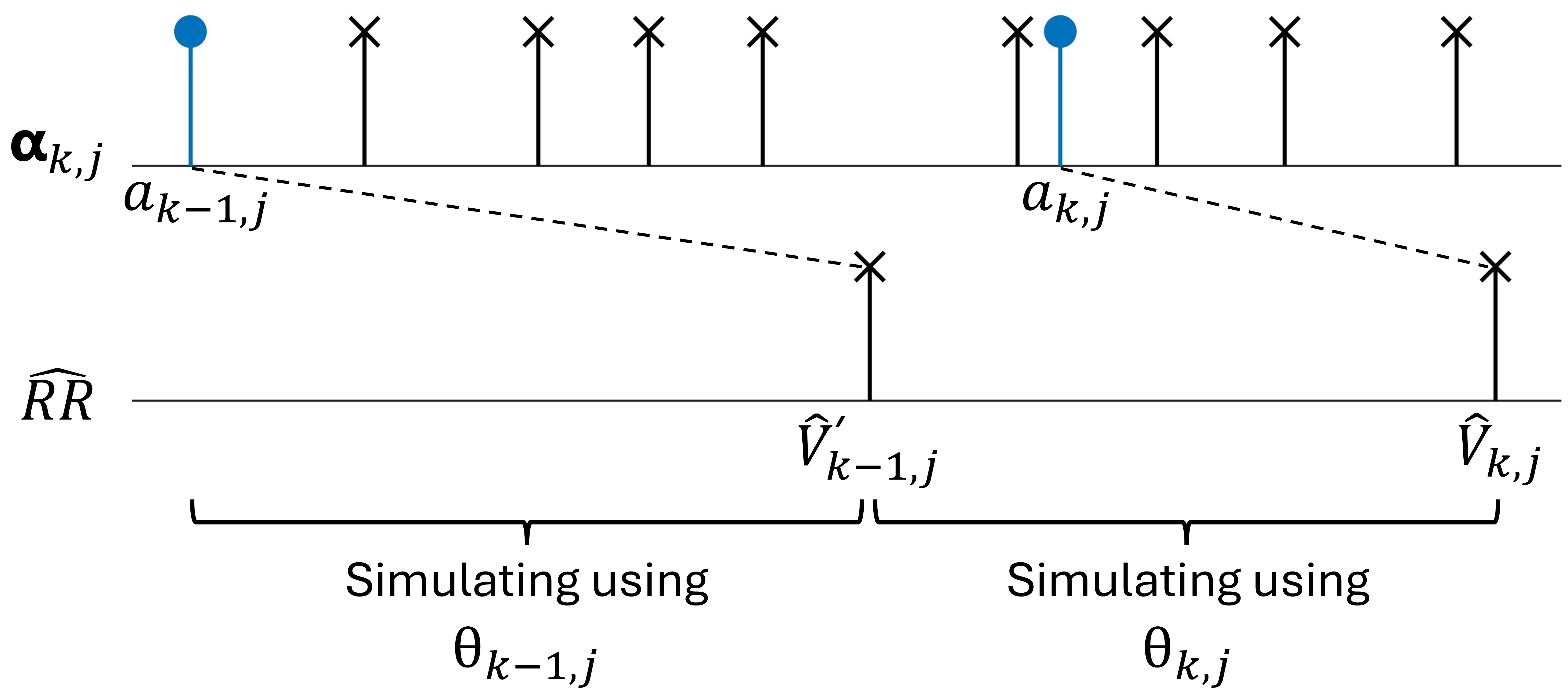}
    \caption{An AA series $\boldsymbol{\alpha}_{k,j}$ and corresponding time series of simulated ventricular activations ($\hat{V}_{k,j}^{'}$ and $\hat{V}_{k,j}$), where $a_{k-1,j}$ leads to $\hat{V}_{k,j}^{'}$. Note that it is not necessarily the first AA impulse after a ventricular activation that leads to the next ventricular activation since impulses may be blocked.}
    \label{Figure2}
\end{figure*}

\subsection{Network model of the AV node} \label{Sec:3_3}
\noindent
Our previously introduced network model of the AV node \cite{karlsson2021non} describes it as two pathways (FP and SP). Each pathway comprises 10 nodes, interconnected with a coupling node at the end, which can transmit impulses to the ventricles (see Figure \ref{Figure1}). Each node corresponds anatomically to a localized section of its respective pathway, while the coupling node represents the Purkinje fibers and Bundle of His \cite{kurian2010anatomy}.\newline
\indent
The AA series extracted from data (see Section \ref{sec:rr_aa}) arrives at the first nodes of the FP and the SP simultaneously. Each node can be refractory (blocking impulses) or non-refractory (transmitting impulses). Transmitted impulses arrive at adjacent nodes with an added conduction delay, and nodes immediately become refractory after transmitting an impulse. The refractory period ($R_i(n)$) and conduction delay ($D_i(n)$) for node $i$ are updated for each incoming impulse $n$ according to Equations \ref{eq:R}, \ref{eq:D}, and \ref{eq:TDI},

\begin{equation} \label{eq:R}
    R_i(n) = R_{min} + \Delta R(1-e^{-\tilde{t}_{i}(n)/\tau_R})
\end{equation}
\begin{equation} \label{eq:D}
    D_i(n) = D_{min} + \Delta D e^{-\tilde{t}_{i}(n)/\tau_D},
\end{equation}
\begin{equation} \label{eq:TDI}
    \tilde{t}_{i}(n) = t_i(n) - ( t_i(n-1) + R_i(n-1) ),
\end{equation}

where $\tilde{t}_i(n)$ is the diastolic interval preceding impulse $n$ and $t_i(n)$ is the arrival time of impulse $n$ at node $i$. When $\tilde{t}_{i}(n) < 0$, the node is in its refractory state and will block incoming impulses. Each pathway has three parameters for the refractory period and three for the conduction delay, totaling 12 model parameters $\boldsymbol{\theta} = [R_{min}^{FP},\ \Delta R^{FP},\ \tau_R^{FP}, R_{min}^{SP},\ \Delta R^{SP},\ \tau_R^{SP}, D_{min}^{FP},\ \Delta D^{FP}$ $,\ \tau_D^{FP}, D_{min}^{SP},\ \Delta D^{SP},\ \tau_D^{SP}]$. The coupling node's refractory period is fixed to the shortest RR interval in the data used for estimation minus 50 ms, and its conduction delay is fixed at 60 ms \cite{karlsson2021non}.\newline
\indent
The model is evaluated using a modified version of Dijkstra’s algorithm \cite{wallman2018characterisation}. Impulses are propagated through the network in an event-based fashion where the impulse with the lowest $t_i(n)$ in a queue ($\boldsymbol{q}(t)$) of impulses is propagated next (or blocked depending on $\tilde{t}_i(n)$). Three types of data are needed to run the model. First, the parameter vector $\boldsymbol{\theta}$ is necessary, corresponding to the properties of the AV node. Second, the queue $\boldsymbol{q}(t)$ is required, where each impulse is represented by a tuple containing arrival time $t_i(n)$ and node index $i$. These impulses can arrive from the atria or from a transmitting node within the model. The atrial activation times from an AA series are placed in $\boldsymbol{q}(t)$ with corresponding node index for the first nodes in the slow and fast pathway. Finally, a vector containing the repolarization times ($\boldsymbol{RT}(t)$) for each of the 21 nodes in the model is needed, corresponding to the end of the diastolic interval ($\tilde{t}_{i}(n)$). The model code and a basic user example can be found at \url{https://github.com/FraunhoferChalmersCentre/AV-node-model}.\newline
\indent The AV node model parameters $\boldsymbol{\theta}$ are assumed to be fixed between two heartbeats and the corresponding AV node conduction properties $\boldsymbol{\phi}$ are estimated in the following way: for a given pathway, e.g. FP, $R^{FP}$ and $D^{FP}$ are estimated as the medians of all of $R_i(n)$ and $D_i(n)$ (Equation \ref{eq:R} and \ref{eq:D}) in the time interval, with $D^{FP}$ multiplied by 10 to account for the cumulative delay of the whole pathway. Values for SP are computed analogously.

\subsection{Parameter estimation}
\noindent
To estimate ${\boldsymbol{\phi}}$ with beat-to-beat resolution, a particle filter is first used to solve the filtering problem -- estimating the current state of the system based on current and past observations -- before a smoothing algorithm is applied to the estimated states to solve the smoothing problem -- estimating the current state of the system based on current, past, and future observations. Moreover, two versions of the particle filter were developed, one designed for RR and AA series extracted from ECG and EGM recordings (EGM-PF), and one designed for AA series generated based on the f-wave frequencies and RR series extracted from the ECG (ECG-PF). \newline

\subsubsection{EGM-particle filter} \label{Sec:PF_known}
\noindent
A basic particle filter can be described by its four phases: initialization, weighting, resampling, and propagation. These phases all affect the particles in the particle filter. In this work, each particle corresponds to a model parameter vector $\boldsymbol{\hat{\theta}}_{k, j}$, where $k$ denotes the RR interval index (also referred to as time step) and $j$ is the particle index. The EGM-PF is initialized by drawing $N = 1,000,000$ particles independently from a twelve-dimensional uniform distribution (ranges found in Supplementary Material S2), where particles with an SP refractory period greater than the FP refractory period or an SP conduction delay less than the FP conduction delay are excluded. Initialization is followed by a weighting phase. This starts by evaluating all $\boldsymbol{\hat{\theta}}_{1, j}$ with the model, using the current repolarization times ($\boldsymbol{RT}_{j}(0) = 0$) and the current queue ($\boldsymbol{q}_{j}(0)$) filled by the AA series extracted from the EGM, until each particle has generated a ventricular activation at time $t = \hat{V}_{k,j}$, i.e. a simulated heartbeat. The resulting $\boldsymbol{q}_{j}(\hat{V}_{k,j})$ and $\boldsymbol{RT}_{j}(\hat{V}_{k,j})$ at the time of each new heartbeat $\hat{V}_{k,j}$ is saved. After all $N$ particles in the filter have been used to simulate heartbeats, each $\hat{V}_{k,j}$ is used together with the time of the true heartbeat ($V_k$) to calculate the weight $w_{k,j}$ of particle $j$. The weight is related to the probability that $\hat{V}_{k,j} - V_k$ was drawn from a normal distribution with zero mean and standard deviation $\sigma_w$, i.e.,

\begin{equation} \label{eq:weight}
    w_{k,j} = {\mathcal{N}}( ~(\hat{V}_{k,j} - V_k)~ | 0, \sigma_w^2),
\end{equation}

\noindent where $\sigma_w$ was set to 30 ms to account for uncertainties in R wave detection. After all $j$ weights have been calculated, they are further normalized by the sum of all weights. The weighting phase is followed by a resampling phase, where new particles ($\boldsymbol{\hat{\theta}}_{k+1, j}$) with corresponding $\boldsymbol{q}_{j}(\hat{V}_{k,j})$ and $\boldsymbol{RT}_{j}(\hat{V}_{k,j})$ are drawn with replacement from $\boldsymbol{\hat{\theta}}_{k,j}$ with probability proportional to their weights, thereby approximating the posterior distribution at time step $k$. In the subsequent propagation phase, each particle is propagated one time step forward by adding normally distributed noise drawn from ${\mathcal{N}}(0, \Sigma)$. Details on $\Sigma$ and the propagation phase are found in Supplementary Material S2. The propagation phase is followed by a new weighting phase. The resampling, propagation, and weighting are repeated sequentially for each time step, from $k = 2$ to $k = K$, where $K$ denotes the last time step. The pseudo-code for the EGM-PF is shown in Algorithm \ref{Code_known}. \newline

\begin{algorithm*}
\caption{EGM-particle filter}
\label{Code_known}
\footnotesize
\begin{algorithmic}
\STATE \hspace{0.2cm} \textbf{Initialization (k = 1):}
\STATE \hspace{0.2cm} \textbf{for} $j = 1$ to $N$ \textbf{do}
\STATE \hspace{0.4cm} Sample $\boldsymbol{\hat{\theta}}_{1, j} \sim U$ with exclusion criteria described in Sec \ref{Sec:PF_known}.
\STATE \hspace{0.4cm} \textbf{Weighting:} 
\STATE \hspace{0.4cm} Simulate $\hat{V}_{1, j}$ by running the model with $\boldsymbol{\hat{\theta}}_{1, j}$, $\boldsymbol{q}_{j}(0)$, and $\boldsymbol{RT}_{j}(0) = \mathbf{0}$.
\STATE \hspace{0.4cm} Save $\boldsymbol{q}_j(\hat{V}_{1,j})$ and $\boldsymbol{RT}_{j}(\hat{V}_{1,j})$ at ventricular activation time $\hat{V}_{1,j}$.
\STATE \hspace{0.4cm} Compute $w_{1, j} = {\mathcal{N}}(~(\hat{V}_{1, j} - V_1)~ | 0, \sigma_w)$ (Eq. ~\ref{eq:weight}).
\STATE \hspace{0.2cm} \textbf{end}
\STATE \hspace{0.2cm} Normalize $w_{1, j} \leftarrow w_{1, j} / \sum_j w_{1, j}$.

\STATE \hspace{0.2cm} \textbf{for} $k = 2$ to $K$ \textbf{do}
\STATE \hspace{0.4cm} \textbf{for} $j = 1$ to $N$ \textbf{do}
\STATE \hspace{0.6cm} \textbf{Resampling:} 
\STATE \hspace{0.6cm} Generate $\hat{\boldsymbol{\theta}}_{k,j}$ by resampling $\boldsymbol{\hat{\theta}}_{k-1,j}$ using $w_{k-1,j}$.

\STATE \hspace{0.6cm} \textbf{Propagation:} 
\STATE \hspace{0.6cm} Sample $\boldsymbol{\hat{\theta}}_{k,j} \sim {\mathcal{N}}(\boldsymbol{\hat{\theta}}_{k-1,j}, \Sigma)$.

\STATE \hspace{0.6cm} \textbf{Weighting:}
\STATE \hspace{0.6cm} Simulate $\hat{V}_{k,j}$ by running the model with $\boldsymbol{\hat{\theta}}_{k,j}$, $\boldsymbol{q}_j(\hat{V}_{k-1,j})$, and $\boldsymbol{RT}_{j}(\hat{V}_{k-1,j})$.
\STATE \hspace{0.6cm} Save $\boldsymbol{q}_j(\hat{V}_{k,j})$ and $\boldsymbol{RT}_{j}(\hat{V}_{k,j})$ at ventricular activation time $\hat{V}_{k,j}$.
\STATE \hspace{0.6cm} Compute $w_{k,j} = {\mathcal{N}}(~(\hat{V}_{k,j} - V_k)~ | 0, \sigma_w)$ (Eq. ~\ref{eq:weight}).
\STATE \hspace{0.4cm} \textbf{end}
\STATE \hspace{0.4cm} Normalize $w_{k,j} \leftarrow w_{k,j} / \sum_j w_{k,j}$.
\STATE \hspace{0.2cm} \textbf{end}
\end{algorithmic}
\end{algorithm*}

\subsubsection{ECG-particle filter}\label{Sec:PF_unknown}
\noindent Similar to the EGM-PF, the ECG-PF is described by its four phases: initialization, weighting, resampling, and propagation. In contrast to the EGM-PF, the atrial activity is not fully known from the ECG. To evaluate several possible AA series, $N = 40,000$ particles are first independently drawn from the same twelve-dimensional uniform distribution as the EGM-PF during initialization, after which $N_{ECG} = 25$ copies of each particle are created ($\boldsymbol{\hat{\theta}}_{k,j}$). Additionally, normally distributed noise drawn from ${\mathcal{N}}(0, \Sigma)$ is added to each particle, with previously defined $\Sigma$ (Supplementary Material S2). This creates $N \cdot N_{ECG}$ unique particles, identical to the number of particles in the EGM-PF. \newline
\indent Each unique particle is evaluated with a different AA series ($\boldsymbol{\alpha}_{k,j}$), with each $\boldsymbol{\alpha}_{k,j}$ generated by a Gaussian random walk, as described in Section \ref{Sec:AA:tilt}. For the first time step, each particle is evaluated by running the model with $\boldsymbol{\hat{\theta}}_{1,j}$ until the first ventricular activation has been simulated ($\hat{V}_{1,j}$). \newline
\indent For the following time steps $k>1$, the AA series ($\boldsymbol{\alpha}$) is generated based on a Gaussian random walk (see Section \ref{Sec:AA:tilt}), with an added impulse at time zero. However, the arrival time of the first impulse is set to the arrival time of the atrial impulse leading to $\hat{V}_{k-1,j}$, denoted $a_{k-1,j}$, so that $\boldsymbol{\alpha}_{k,j} = \boldsymbol{\alpha} + a_{k-1,j}$. Moreover, each particle is evaluated by running the model until \textit{two} ventricular activations have been simulated ($\hat{V}_{k-1,j}^{'}$ and $\hat{V}_{k,j}$). Before the time $\hat{V}_{k-1,j}$, the parameter vector $\boldsymbol{\hat{\theta}}_{k-1, j}$ is used to evaluate the model, whereas $\boldsymbol{\hat{\theta}}_{k, j}$ is used after $\hat{V}_{k-1,j}$, as illustrated in Figure \ref{Figure2}. Around 98\% of the time, $\hat{V}_{k-1,j}^{'}$ is equal to $\hat{V}_{k-1,j}$, and particles where $\hat{V}_{k-1,j}^{'}$ differ from $\hat{V}_{k-1,j}$ are excluded. Consequently, when running the particle filter, $\hat{V}_{k-1,j}^{'}$ is equivalent to $\hat{V}_{k-1,j}$. This leads to impulses in $\boldsymbol{\alpha}_{k-1,j}$ arriving after $a_{k-1,j}$ not affecting $\hat{V}_{k-1,j}$ in the particle filter and thus do not affect the probability of being selected for the previous resampling. Therefore, these impulses should not affect $\hat{V}_{k,j}$. To ensure this, re-running the previous time step in this manner was performed.\newline
%
\indent After $\hat{V}_{k,j}$ has been generated by the model, values for $\hat{V}_{k,j}$, $\boldsymbol{q}_{j}(\hat{V}_{k,j})$, and $\boldsymbol{RT}_{j}(\hat{V}_{k,j})$ are saved for each particle. As for the EGM-PF, each $\hat{V}_{k,j}$ is used together with a corresponding measured value $V_k$ in Equation \ref{eq:weight} to calculate the weight $w_{k,j}$. \newline
\indent In the resampling phase, $N$ new particles with corresponding $\boldsymbol{q}_{j}(\boldsymbol{a}_{k,j})$ and $\boldsymbol{RT}_{j}(\boldsymbol{a}_{k,j})$ are drawn with replacement from $\boldsymbol{\hat{\theta}}_{k,j}$ based on their weights, before $N_{ECG} = 25$ copies of each particle are created. Normally distributed noise drawn from ${\mathcal{N}}(0, \Sigma)$ is added to each of the copied particles, which functions as the propagation phase, thereby creating $\boldsymbol{\hat{\theta}}_{k+1,j}$.\newline
\indent The propagation phase is followed by a new weighting phase before the resampling, propagation, and weighting are repeated sequentially for each time step, from $k = 2$ to $k = K$. The pseudo-code for the ECG-PF is shown in Algorithm \ref{Code_unknown}. In addition, Matlab code with a usage example can be found at \url{https://github.com/FraunhoferChalmersCentre/AV-node-model}.\newline

\begin{algorithm*}
\caption{ECG-particle filter. Differences from the EGM-PF are marked with '*'.}
\label{Code_unknown}
\footnotesize
\begin{algorithmic}
\STATE \hspace{0.2cm} \textbf{Initialization (k = 1):}
\STATE \hspace{0.2cm} \textbf{for} $j = 1$ to $N$ \textbf{do}
\STATE \hspace{0.4cm} Sample $N$ $\boldsymbol{\hat{\theta}} \sim U$ with exclusion criteria described in Sec \ref{Sec:PF_known}.
\STATE *\hspace{0.28cm} Copy $\boldsymbol{\hat{\theta}}$ $N_{ECG}$ times to generate $\boldsymbol{\hat{\theta}}_{1, j}$
\STATE *\hspace{0.28cm} Sample $\boldsymbol{\hat{\theta}}_{1,j} \sim {\mathcal{N}}(\boldsymbol{\hat{\theta}}_{1, j}, \Sigma)$
\STATE \hspace{0.4cm} \textbf{Weighting:}
\STATE *\hspace{0.28cm} Simulate $\hat{V}_{1, j}$ by running the model with $\boldsymbol{\hat{\theta}}_{1, j}$,  $\boldsymbol{q}_{j}(0) = \boldsymbol{\alpha}_{1,j}$, and $\boldsymbol{RT}_{j}(0) = \mathbf{0}$.
\STATE \hspace{0.4cm} Save $\boldsymbol{q}_j(a_{1,j})$ and $\boldsymbol{RT}_{j}(a_{1,j})$ at ventricular activation time $\hat{V}_{1,j}$.
\STATE \hspace{0.4cm} Compute $w_{1, j} = {\mathcal{N}}(~(\hat{V}_{1, j} - V_1)~ | 0, \sigma_w)$ (Eq. ~\ref{eq:weight}).
\STATE \hspace{0.2cm} \textbf{end}
\STATE \hspace{0.2cm} Normalize $w_{1, j} \leftarrow w_{1, j} / \sum_j w_{1, j}$.

\STATE \hspace{0.2cm} \textbf{for} $k = 2$ to $K$ \textbf{do}
\STATE \hspace{0.4cm} \textbf{for} $j = 1$ to $N$ \textbf{do}
\STATE \hspace{0.6cm} \textbf{Resampling:} 
\STATE \hspace{0.6cm} Generate $N$ $\hat{\boldsymbol{\theta}}$ by resampling $\boldsymbol{\hat{\theta}}_{k-1,j}$ using $w_{k-1,j}$.

\STATE *\hspace{0.48cm} Copy $\boldsymbol{\hat{\theta}}$ $N_{ECG}$ times to generate $\boldsymbol{\hat{\theta}}_{k, j}$

\STATE \hspace{0.6cm} \textbf{Propagation:} 
\STATE \hspace{0.6cm} Sample $\boldsymbol{\hat{\theta}}_{k,j} \sim {\mathcal{N}}(\boldsymbol{\hat{\theta}}_{k, j}, \Sigma)$.

\STATE \hspace{0.6cm} \textbf{Weighting:} 
\STATE *\hspace{0.48cm} Simulate $\hat{V}_{k,j}$ by running the model as described in \ref{Sec:PF_unknown} with $\boldsymbol{\hat{\theta}}_{k-1,j}$, $\boldsymbol{\hat{\theta}}_{k,j}$, $\boldsymbol{q}_j(\boldsymbol{a}_{k-1,j})$, $\boldsymbol{RT}_{j}(\boldsymbol{a}_{k-1,j})$, and $\boldsymbol{\alpha}_{k,j}$ generated as described in \ref{Sec:AA:tilt}.

\STATE \hspace{0.6cm} Compute $w_{k,j} = {\mathcal{N}}(~(\hat{V}_{k,j} - V_k)~ | 0, \sigma_w)$ (Eq. ~\ref{eq:weight}).
\STATE \hspace{0.4cm} \textbf{end}
\STATE \hspace{0.4cm} Normalize $w_{k,j} \leftarrow w_{k,j} / \sum_j w_{k,j}$.
\STATE \hspace{0.2cm} \textbf{end}
\end{algorithmic}
\end{algorithm*}

\subsubsection{Smoothing algorithm} \label{Sec:smooth}
\noindent
The combined particle filter and smoothing algorithm utilized in this work is commonly referred to as the forward filtering backward sampling algorithm \cite{chopin2020introduction}. The smoothing algorithm is applied after either the EGM-PF or the ECG-PF, and functions as the backward sampling step. \newline
\indent Starting at the last time step $k=K$, one of the $j$ particles is selected with probability proportional to $w_{K,j}$ and denoted $\boldsymbol{x}(K)$, where $\boldsymbol{x}$ is a vector of indices. The algorithm continues iteratively from time step $k = K-1$ to $k = 1$. The weights are updated based on the likelihood that $\boldsymbol{\hat{\theta}}_{k,j}$ originates from the selected particle at the time step of the previous iteration $\boldsymbol{\hat{\theta}}_{k+1, \boldsymbol{x}(k+1)}$, with previously defined $\Sigma$ (see Supplementary Material S2), according to Equation \ref{eq:smooth}. 

\begin{equation} \label{eq:smooth}
    \hat{w}_{k,j} = w_{k,j}{\mathcal{N}}( \boldsymbol{\hat{\theta}}_{k+1,\boldsymbol{x}(k+1)} | \boldsymbol{\hat{\theta}}_{k,j}, \Sigma).
\end{equation}

A new particle $j$ is selected with probability proportional to $\hat{w}_{k,j}$ and assigned to $\boldsymbol{x}(k)$. After completion, the vector $\boldsymbol{x}$ contains one trajectory of indices corresponding to parameters $\boldsymbol{\hat{\theta}}_{k,x(k)}$ sampled from the smoothing probability density function. Running the smoothing algorithm $M = 20,000$ times generates $M$ trajectories ($\boldsymbol{x}_m$) of $\boldsymbol{\hat{\theta}}$, all sampled from the smoothing probability density function. The pseudo-code for the smoothing algorithm is shown in Algorithm \ref{Smoothing}. Since each simulated heartbeat is associated with a $\hat{\boldsymbol{\phi}}$ (see Section \ref{Sec:3_3}), the $M$ trajectories $\boldsymbol{x}_m(k)$ also yield $M$ trajectories of $\hat{\boldsymbol{\phi}}$ sampled from the smoothing probability density function. These are used as estimates of the posterior distribution of the AV node conduction delays and refractory periods and denoted $\tilde{\boldsymbol{\phi}}_{m}(k)$. \newline

\vspace{-4 mm}
\begin{algorithm*}
\caption{Smoothing algorithm}
\label{Smoothing}
\footnotesize
\begin{algorithmic} \label{Code_smooth}
\STATE \hspace{0.2cm} \textbf{Initialization (k = K):}
\STATE \hspace{0.2cm} \textbf{Sample $\boldsymbol{x}(K) \sim w_{K,j}$}
\STATE \hspace{0.2cm} \textbf{for} $k = K-1$ to $1$ \textbf{do}
\STATE \hspace{0.4cm} \textbf{for} $j = 1$ to $N$ \textbf{do}
\STATE \hspace{0.6cm} \textbf{Update weights:} $\hat{w}_{k,j} \leftarrow w_{k,j} {\mathcal{N}}(\boldsymbol{\hat{\theta}}_{k+1,\boldsymbol{x}(k+1)} | \boldsymbol{\hat{\theta}}_{k,j}, \Sigma)$ (Eq. ~\ref{eq:smooth})
\STATE \hspace{0.4cm} \textbf{end}
\STATE \hspace{0.4cm} \textbf{Sample:} $\boldsymbol{x}(k) \sim \hat{w}_{k,j}$
\STATE \hspace{0.2cm} \textbf{end}
\end{algorithmic}
\end{algorithm*}

\begin{figure*}[b]
    \centering
        \includegraphics[width=18cm]{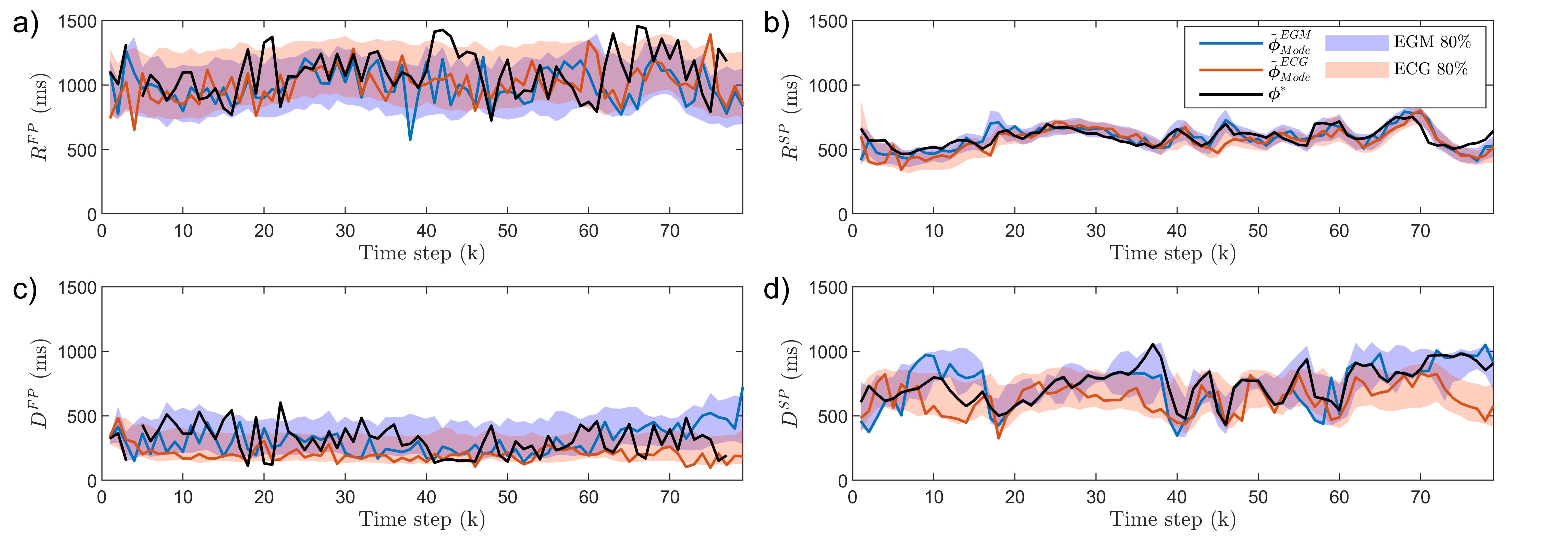}
    \caption{AV node estimates $\tilde{\boldsymbol{\phi}}^{EGM}_{m}(k)$ (blue) and $\tilde{\boldsymbol{\phi}}^{ECG}_{m}(k)$ (red) obtained based on simulated data and corresponding ground truth $\boldsymbol{\phi}^*(k)$ (black): (a) $R^{FP}$, (b) $R^{SP}$, (c) $D^{FP}$, (d) $D^{SP}$. For comparison with Table \ref{tab:res:KS-mode}, EGM and ECG $\overline{l^1}$ are 183 ms and 160 ms in $R^{FP}$, 46 ms and 58 ms in $R^{FP}$, 114 ms and 127 ms in $R^{FP}$, and 76 ms and 138 ms in $R^{FP}$. The modes $\tilde{\boldsymbol{\phi}}^{EGM}_{Mode}(k)$ and $\tilde{\boldsymbol{\phi}}^{ECG}_{Mode}(k)$ are shown as solid lines and the 80\% credibility region as shaded background. The 80\% credibility region is used over the 95\% for clarity of visualization.}
    \label{Figure3}
\end{figure*}

\subsection{Evaluation of particle filters} \label{Sec:evo:sim}
\noindent To evaluate the accuracy of the proposed methodology, estimates obtained using the EGM-PF ($\tilde{\boldsymbol{\phi}}^{EGM}_{m}(k)$) and ECG-PF ($\tilde{\boldsymbol{\phi}}^{ECG}_{m}(k)$), respectively, are compared with the corresponding ground truth from the simulations. Recordings from the iafdb (see Section \ref{Sec:iafdb}) are used to create realistic simulated data, as described in Supplementary Section S1. This results in 50 model parameter trends $\boldsymbol{\theta}^*(k)$ with associated AA series and f-waves, needed for the EGM-PF and ECG-PF, respectively. Each parameter trend $\boldsymbol{\theta}^*(k)$ with corresponding AA series was used to simulate an RR series of one minute duration with associated ground truth AV node conduction property trends ($\boldsymbol{\phi}^*(k)$). \newline
\indent To compute the most probable value at each time step $k$ for each AV node property, the mode of the estimated trends $\tilde{\boldsymbol{\phi}}^{EGM}_{m}(k)$ and $\tilde{\boldsymbol{\phi}}^{ECG}_{m}(k)$ are computed and compared to the mode of the ground truth $\boldsymbol{\phi}^*(k)$. The mode is obtained independently for each time step $k$ by sorting the $M$ values into histogram bins with 5 ms width before identifying the center of the histogram bin with the highest count, resulting in $\tilde{\boldsymbol{\phi}}^{EGM}_{Mode}(k)$ and $\tilde{\boldsymbol{\phi}}^{ECG}_{Mode}(k)$. The bin width of 5 ms was chosen to strike a balance between temporal resolution and robustness to noise in the estimated distributions. The average $l^1$-norm ($\overline{l^1}$), between $\boldsymbol{\phi}^*(k)$ and $\tilde{\boldsymbol{\phi}}^{EGM}_{Mode}(k)$ and $\tilde{\boldsymbol{\phi}}^{ECG}_{Mode}(k)$, respectively, for each AV node property in each simulation is used to quantify how well the most probable estimate aligns with the ground truth. To facilitate comparison between the AV node conduction properties, $\overline{l^1}$ is also normalized to a percentage by dividing by the range between the highest and lowest values for each property in the simulations $\boldsymbol{\phi}^*(k)$, denoted $\boldsymbol{r}$ (values found in Supplementary Material S1). In addition, $\tilde{\boldsymbol{\phi}}^{EGM}_{Mode}(k)$, $\tilde{\boldsymbol{\phi}}^{ECG}_{Mode}(k)$, and $\boldsymbol{\phi}^*(k)$ are averaged over each one-minute simulation, and the $l^1$-norm between the averages, denoted $\overline{l^1_{min}}$, is used to evaluate how well the proposed methodology performs with a temporal resolution of one minute. \newline
\indent Furthermore, the percentage of heartbeats for which the 95\% credibility region covered $\boldsymbol{\phi}^*(k)$, denoted $CR_{95}(k)$, is calculated by finding the values at the 2.5th and 97.5th percentiles of $\tilde{\boldsymbol{\phi}}^{EGM}_{m}(k)$ and $\tilde{\boldsymbol{\phi}}^{ECG}_{m}(k)$ and evaluating how often $\boldsymbol{\phi}^*(k)$ lies in between. In theory, this should converge towards 95\% as the number of samples grows. \newline

\subsection{Analysis of iafdb data} \label{Sec:evo:iafdb}
\noindent EGM recordings from the tip of the tricuspid valve annulus with synchronized ECG recordings, as described in Section \ref{Sec:iafdb}, are analyzed to compare the concordance between the resulting estimates from the EGM-PF and the ECG-PF. The $\tilde{\boldsymbol{\phi}}^{EGM}_{m}(k)$ and $\tilde{\boldsymbol{\phi}}^{ECG}_{m}(k)$ are estimated using the two particle filters and the smoothing algorithm before $\tilde{\boldsymbol{\phi}}^{EGM}_{Mode}(k)$ and $\tilde{\boldsymbol{\phi}}^{ECG}_{Mode}(k)$ are calculated as described in Section \ref{Sec:evo:sim}. The concordance between $\tilde{\boldsymbol{\phi}}^{EGM}_{Mode}(k)$ and $\tilde{\boldsymbol{\phi}}^{ECG}_{Mode}(k)$ is visualized using the Bland–Altman plot.

\subsection{Analysis of tilt test data}
\noindent
ECG recordings from a tilt test protocol, as described in \ref{Sec:tilt}, are analyzed to evaluate the method's ability to quantify expected changes in AV node characteristics. First, $\tilde{\boldsymbol{\phi}}^{ECG}_{m}(k)$ is estimated using the ECG-PF and smoothing algorithm, before $\tilde{\boldsymbol{\phi}}^{ECG}_{Mode}(k)$ is calculated as described in Section \ref{Sec:evo:sim}. Further, for each patient, $\tilde{\boldsymbol{\phi}}^{ECG}_{Mode}(k)$ is averaged over each tilt phase to obtain $\overline{\boldsymbol{\phi}}^{Supine}$, $\overline{\boldsymbol{\phi}}^{HDT}$, and $\overline{\boldsymbol{\phi}}^{HUT}$, which are used to analyze changes between phases. \newline

\subsection{Statistical analysis}
\noindent The paired one-sided Wilcoxon signed rank test is used to quantify significant increase or decrease in this study, since the data do not generally follow a normal distribution according to the Shapiro-Wilk test ($p < 0.05$). This includes the paired significant test between $\overline{l^1}$, $\overline{l^1_{min}}$, $\overline{l^1}/\boldsymbol{r}$, and $\overline{l^1_{min}}/\boldsymbol{r}$ obtained using $\tilde{\boldsymbol{\phi}}^{EGM}_{Mode}(k)$ and $\tilde{\boldsymbol{\phi}}^{ECG}_{Mode}(k)$, respectively, as well as the paired difference between $\overline{\boldsymbol{\phi}}^{Supine}$ and $\overline{\boldsymbol{\phi}}^{HDT}$ and between $\overline{\boldsymbol{\phi}}^{Supine}$ and $\overline{\boldsymbol{\phi}}^{HUT}$.

\begin{figure*}[b]
    \centering
        \includegraphics[width=18cm]{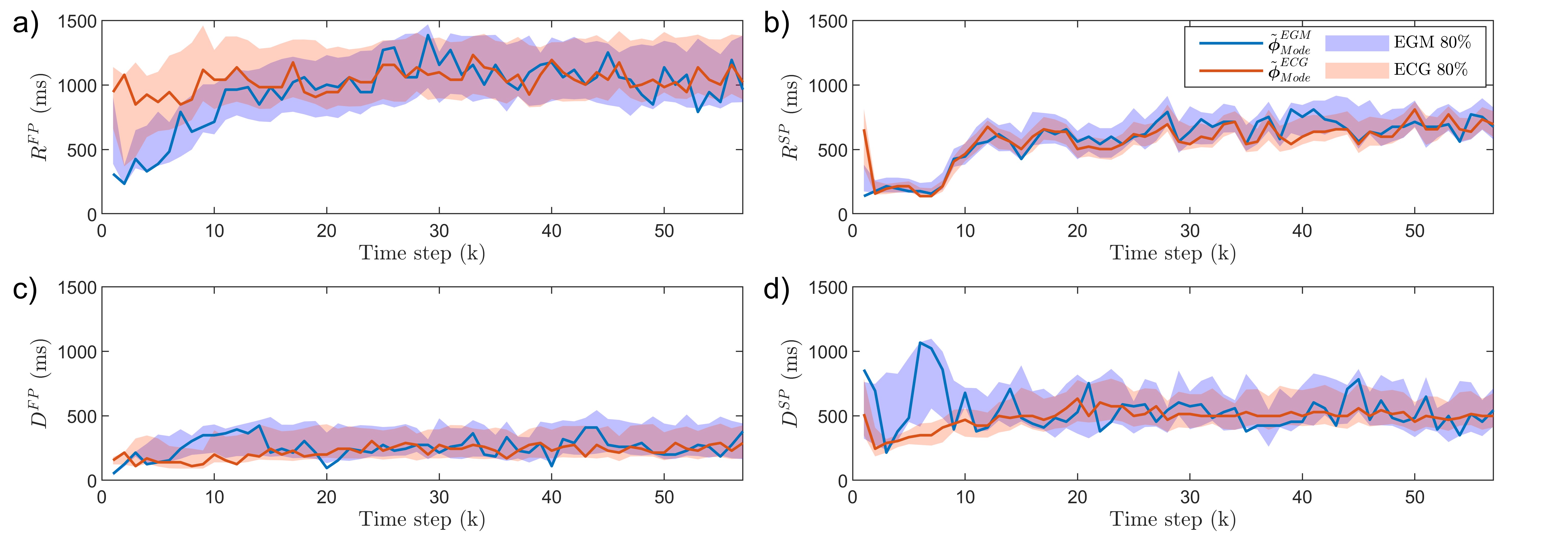}
    \caption{The AV node estimates $\tilde{\boldsymbol{\phi}}^{EGM}_{m}(k)$ (blue) and $\tilde{\boldsymbol{\phi}}^{ECG}_{m}(k)$ (red) obtained for Patient 6 in the iafdb database: (a) $R^{FP}$, (b) $R^{SP}$, (c) $D^{FP}$, (d) $D^{SP}$. The modes $\tilde{\boldsymbol{\phi}}^{EGM}_{Mode}(k)$ and $\tilde{\boldsymbol{\phi}}^{ECG}_{Mode}(k)$ are shown as solid lines and the 80\% credibility region as shaded background. }
    \label{Figure4}
\end{figure*}

\vspace{-4pt}
\section{Results} \label{Sec:result}
To recapitulate, three datasets were used in this study. The simulated data (see Section \ref{Sec:evo:sim}) are used to evaluate the estimation accuracy for the EGM and ECG-based methods. The iafdb recordings (see Section \ref{Sec:iafdb}) are used to compare the estimates obtained from synchronized EGM and ECG recordings. The tilt recordings (see Section \ref{Sec:tilt}) are used to evaluate the method's ability to quantify expected changes in AV node characteristics. \newline
\indent The computation time (performed on a desktop computer with a Ryzen 9 5900X CPU, using the twelve cores in parallel) to obtain posterior distributions $\tilde{\boldsymbol{\phi}}^{EGM}_{m}(k)$ and $\tilde{\boldsymbol{\phi}}^{ECG}_{m}(k)$ was on average 3.7 and 17.5 minutes per minute of data, respectively.

\subsection{Evaluation of particle filters} \label{Sec:res:sim}
\noindent An example of $\tilde{\boldsymbol{\phi}}^{EGM}_{m}(k)$ and $\tilde{\boldsymbol{\phi}}^{ECG}_{m}(k)$, estimated from simulated data, with corresponding ground truth $\boldsymbol{\phi}^*(k)$, is displayed in Figure \ref{Figure3}. The shown example displays typical patterns, such as the estimates of $R^{SP}$ having a more narrow credibility region compared to the estimates of $R^{FP}$, $D^{FP}$, and, $D^{SP}$, as well as more accurate tracking of changes for $D^{SP}$ estimates (Figure \ref{Figure3} d) obtained from EGM recordings compared to ECG recordings. Moreover, neither PF-EGM nor PF-ECG managed to capture fast changes in $\boldsymbol{\phi}^*(k)$ for $R^{FP}$ and $D^{FP}$ (Figure \ref{Figure3} a and c). \newline
\indent The results from the analysis of all simulated data are summarized in Table \ref{tab:res:GT}. As expected, the estimation error $\overline{l^1}$ in $R^{SP}$, $D^{FP}$, and $D^{SP}$ is significantly lower for EGM compared to ECG. However, the opposite is true for $R^{FP}$. The estimate of $R^{SP}$ has the closest match to the ground truth independently of the type of recording used, with an EGM $\overline{l^1}/\boldsymbol{r}$ of 4.48\% and ECG $\overline{l^1}/\boldsymbol{r}$ of 6.33\%. Further, the estimation error of $D^{SP}$ increases the most between EGM and ECG recordings, as seen in an almost doubling of $\overline{l^1}/{\boldsymbol{r}}$, from 8.97\% to 16\%. Not surprisingly, the accuracy increases when compressing the one-minute trend into a single estimate, as seen by the decrease in $\overline{l^1_{min}}$ compared to $\overline{l^1}$ for ECG and EGM for all AV node conduction properties. Additionally, $CR_{95}$ is typically slightly above 95\% for all AV node conduction properties, indicating that the uncertainty bounds produced are conservative if not exact.

\setlength{\tabcolsep}{0.3em} 
{\renewcommand{\arraystretch}{1.25}
\begin{table}[h!] \centering
\caption{The mean $\pm$ standard deviation of $\overline{l^1}$ and $\overline{l^1_{min}}$ between simulated patients, using the EGM-PF and ECG-PF. \dag~ indicate a significant decrease and $^\ddag$ indicates a significant increase for ECG compared to EGM ($p < 0.05$).}
\resizebox{8.75cm}{!}{%
\begin{tabular}{c c c c c }
\toprule 
{\scriptsize  } & {\scriptsize  $R^{FP}$} & {\scriptsize  $R^{SP}$} & {\scriptsize  $D^{FP}$} & {\scriptsize  $D^{SP}$}   \\
\cmidrule(rl){2-2}
\cmidrule(rl){3-3}
\cmidrule(rl){4-4}
\cmidrule(rl){5-5}
EGM $\overline{l^1}$ (ms) & 174 $\pm$ 16.5 & 51.4 $\pm$ 11.5 & 105 $\pm$ 8.7 & 99.5 $\pm$ 23.6 \\
ECG $\overline{l^1}$ (ms) & 169 $\pm$ 13.7$^\dag$ & 66.5 $\pm$ 9.78$^\ddag$ & 131 $\pm$ 13.3$^\ddag$ & 178 $\pm$ 28.1$^\ddag$ \\
EGM $\overline{l^1_{min}}$ (ms) &  141 $\pm$ 35.4 & 16.2 $\pm$ 18.4 & 24.4 $\pm$ 19.5 & 44.3 $\pm$ 33.4 \\
ECG $\overline{l^1_{min}}$ (ms) &  92.3 $\pm$ 52.3$^\dag$ & 23.6 $\pm$ 17.5$^\ddag$ & 89.6 $\pm$ 33.8$^\ddag$ & 152 $\pm$ 65.5$^\ddag$ \\
EGM $\overline{l^1}/\boldsymbol{r}$ (\%) & 12 $\pm$ 1.14 & 4.89 $\pm$ 1.1 & 11 $\pm$ 0.916 & 8.97 $\pm$ 2.13 \\
ECG $\overline{l^1}/\boldsymbol{r}$ (\%) & 11.7 $\pm$ 0.943$^\dag$ & 6.33 $\pm$ 0.931$^\ddag$ & 13.8 $\pm$ 1.4$^\ddag$ & 16 $\pm$ 2.53$^\ddag$ \\
EGM $\overline{l^1_{min}}/\boldsymbol{r}$ (\%) & 9.69 $\pm$ 2.44 & 1.54 $\pm$ 1.75 & 2.57 $\pm$ 2.05 & 3.99 $\pm$ 3.01 \\
ECG $\overline{l^1_{min}}/\boldsymbol{r}$ (\%) & 6.36 $\pm$ 3.60$^\dag$ & 2.24 $\pm$ 1.67$^\ddag$ & 9.43 $\pm$ 3.56$^\ddag$ & 13.7 $\pm$ 5.90$^\ddag$ \\
EGM $CR_{95}$ (\%) & 99.5 $\pm$ 1.1 & 99.4 $\pm$ 1.39 & 98.9 $\pm$ 1.67 & 98.8 $\pm$ 2.38 \\
ECG $CR_{95}$ (\%) & 99.8 $\pm$ 0.543 & 96.4 $\pm$ 3.72 & 98.5 $\pm$ 1.97 & 93 $\pm$ 6.48
\\\bottomrule
\end{tabular} }
\label{tab:res:GT}
\end{table}}

\subsection{Analysis of iafdb data} \label{Sec:res:iafdb}
\begin{figure*}[b]
    \centering
        \includegraphics[width=18cm]{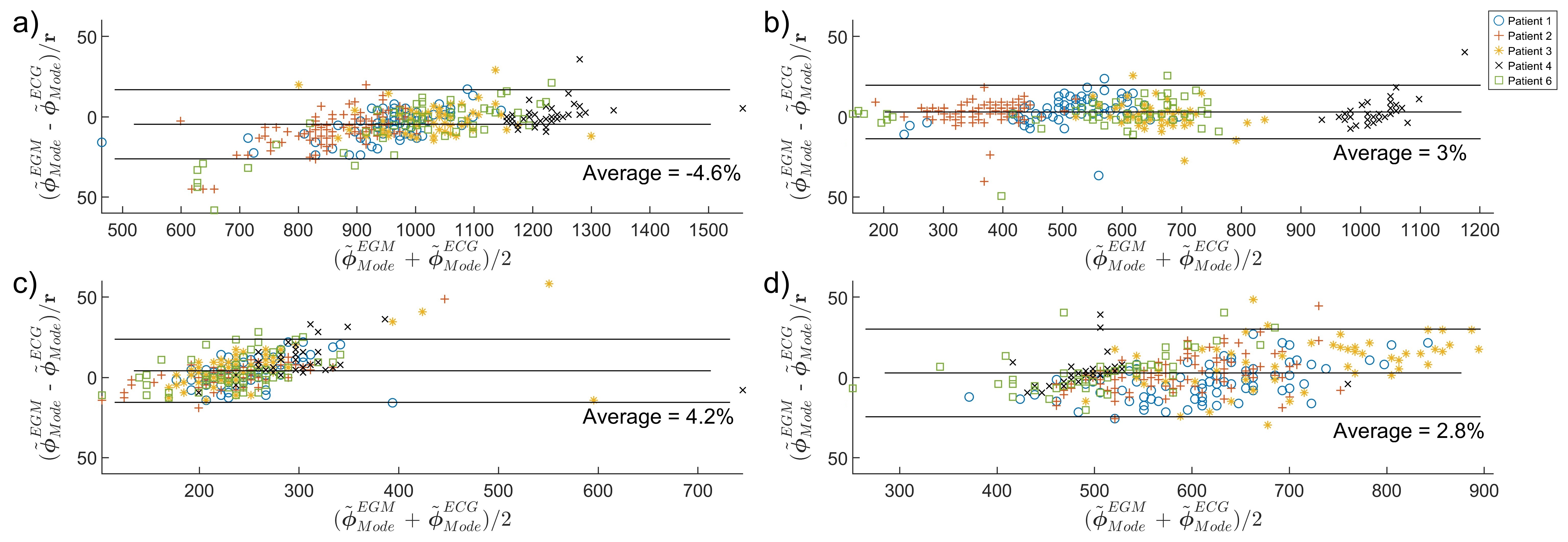}
    \caption{Bland-Altman plot comparing the concordance between ECG-PF and EGM-PF estimates of (a) $R^{FP}$, (b) $R^{SP}$, (c) $D^{FP}$, and (d) $D^{SP}$ obtained from Patient 1 (blue 'o'), Patient 2 (red '+'), Patient 3 (yellow '*'), Patient 4 (black 'x'), and Patient 6 (green '$\square$'). }
    \label{FigureX}
\end{figure*}

\noindent An example of $\tilde{\boldsymbol{\phi}}^{EGM}_{m}(k)$ and $\tilde{\boldsymbol{\phi}}^{ECG}_{m}(k)$ estimated from synchronized EGM and ECG recordings from one patient is shown in Figure \ref{Figure4}. Similar to the results from the analysis of simulated data, the largest difference between using EGM and ECG recordings is seen in $D^{SP}$ (Figure \ref{Figure4} d). Notably, the difference between $\tilde{\boldsymbol{\phi}}^{EGM}_{m}(k)$ and $\tilde{\boldsymbol{\phi}}^{ECG}_{m}(k)$ is more prominent for the first heartbeats, as seen clearly in the $R^{FP}$ (Figure \ref{Figure4} a). This is likely an artifact from the particle filter, which tends to give a higher uncertainty for early time steps. \newline
\indent The results of $\tilde{\boldsymbol{\phi}}^{EGM}_{Mode}(k)$ and $\tilde{\boldsymbol{\phi}}^{ECG}_{Mode}(k)$ for all patient is summarized in Table \ref{tab:res:KS-mode}, and the Bland–Altman plots of $\tilde{\boldsymbol{\phi}}^{EGM}_{Mode}(k)$ and $\tilde{\boldsymbol{\phi}}^{ECG}_{Mode}(k)$ are presented in Figure \ref{FigureX}. On average, $R^{SP}$, $D^{FP}$, and $D^{SP}$ estimated from ECG are slightly lower than the corresponding estimates obtained from EGM, whereas the opposite is true for $R^{FP}$. However, the average difference is less than 5\% for all AV node conduction properties. Differences between $\tilde{\boldsymbol{\phi}}^{EGM}_{Mode}(k)$ and $\tilde{\boldsymbol{\phi}}^{ECG}_{Mode}(k)$ do not seem to be patient-specific. \newline

\setlength{\tabcolsep}{0.3em} 
{\renewcommand{\arraystretch}{1.25}
\begin{table}[h!] \centering
\caption{The mean $\pm$ standard deviation of $\tilde{\boldsymbol{\phi}}^{EGM}_{Mode}(k)$ and $\tilde{\boldsymbol{\phi}}^{ECG}_{Mode}(k)$ computed over all time steps $k$ for all patients in the iafdb.}
\resizebox{8.75cm}{!}{%
\centering
\begin{tabular}{c c c c c c }
\toprule 
{\scriptsize Patient } & {\scriptsize Data } & {\scriptsize  $R^{FP}$ (ms)} & {\scriptsize  $R^{SP}$ (ms)} & {\scriptsize  $D^{FP}$ (ms)} & {\scriptsize  $D^{SP}$ (ms)}   \\
\cmidrule(rl){2-2} \cmidrule(rl){3-3} \cmidrule(rl){4-4} \cmidrule(rl){5-5} \cmidrule(rl){6-6} \cmidrule(rl){1-1}
1 & EGM & 917  $\pm$ 141 & 535 $\pm$ 104 & 257 $\pm$ 67.8 & 593 $\pm$ 135 \\
1 & ECG & 993  $\pm$ 76.3 & 492 $\pm$ 81.6 & 231 $\pm$ 42.5 & 627 $\pm$ 78.9 \\
2 & EGM & 818  $\pm$ 149 & 392 $\pm$ 69.9 & 240 $\pm$ 77.5 & 600 $\pm$ 132 \\ 
2 & ECG & 926  $\pm$ 74.8 & 355 $\pm$ 61.3 & 221 $\pm$ 34.4 & 574 $\pm$ 68.2 \\  
3 & EGM & 1015 $\pm$ 108 & 663 $\pm$ 59.7 & 283 $\pm$ 120 & 761 $\pm$ 164 \\  
3 & ECG & 1037 $\pm$ 101 & 656 $\pm$ 79 & 219 $\pm$ 68.8 & 655 $\pm$ 104 \\ 
4 & EGM & 1240 $\pm$ 102 & 1058 $\pm$ 113 & 355 $\pm$ 92.3 & 507 $\pm$ 81.4 \\  
4 & ECG & 1217 $\pm$ 72.2 & 1004 $\pm$ 37.7 & 259 $\pm$ 93.7 & 478 $\pm$ 67.7 \\  
6 & EGM & 952  $\pm$ 251 & 576 $\pm$ 183 & 253 $\pm$ 82 & 538 $\pm$ 159 \\  
6 & ECG & 1035 $\pm$ 89.6 & 559 $\pm$ 161 & 221 $\pm$ 50.8 & 485 $\pm$ 72.9 
\\\bottomrule
\end{tabular} }
\label{tab:res:KS-mode}
\end{table}}

\subsection{Analysis of tilt test data} \label{Sec:res:tilt}
\noindent
An example of $\tilde{\boldsymbol{\phi}}^{ECG}_{m}(k)$ estimated from one patient during the tilt test is shown in Figure \ref{FigureY}. Notably, transient changes in $R^{SP}$ and $D^{SP}$ can be seen in response to HDT and HUT. \newline
\indent The resulting average estimates for each patient and each phase, $\overline{\boldsymbol{\phi}}^{Supine}$, $\overline{\boldsymbol{\phi}}^{HDT}$, and $\overline{\boldsymbol{\phi}}^{HUT}$ are summarized in Table \ref{tab:res:tilt}. A significant decrease between $\overline{\boldsymbol{\phi}}^{Supine}$ and $\overline{\boldsymbol{\phi}}^{HUT}$ for $R^{FP}$ ($p < 0.05$, 16 of 21 patients), $R^{SP}$ ($p < 0.001$, 18 of 21 patients), and $D^{FP}$ ($p < 0.01$, 16 of 21 patients) could be seen, with no significant difference in $D^{SP}$. Moreover, no significant difference between $\overline{\boldsymbol{\phi}}^{Supine}$ and $\overline{\boldsymbol{\phi}}^{HDT}$ was seen. \newline
\setlength{\tabcolsep}{0.3em} 
{\renewcommand{\arraystretch}{1.25}
\begin{table}[t] \centering
\caption{The population mean $\pm$ standard deviation of the average $\tilde{\boldsymbol{\phi}}^{ECG}_{Mode}(k)$ for each tilt position for the patient in the tilt test, where \dag~indicate a significant decrease ($p < 0.05$) compared with supine position. }
\resizebox{8.75cm}{!}{%
\centering
\begin{tabular}{c c c c c }
\toprule 
{\scriptsize  } & {\scriptsize  $R^{FP}$} & {\scriptsize  $R^{SP}$} & {\scriptsize  $D^{FP}$} & {\scriptsize  $D^{SP}$}   \\
\cmidrule(rl){2-2}
\cmidrule(rl){3-3}
\cmidrule(rl){4-4}
\cmidrule(rl){5-5}
\cmidrule(rl){1-1}
$\overline{\boldsymbol{\phi}}^{Supine}$ (ms) & 926.2 $\pm$ 26.2 & 349.4 $\pm$ 68.8 & 203.5 $\pm$ 13.7 & 523.8 $\pm$ 11.3 \\  
$\overline{\boldsymbol{\phi}}^{HDT}$ (ms)    & 925.5 $\pm$ 24.1 & 338.7 $\pm$ 64.8 & 202.0 $\pm$ 12.4 & 528.8 $\pm$ 12.8 \\   
$\overline{\boldsymbol{\phi}}^{HUT}$ (ms)    & 921.5 $\pm$ 25.3$^\dag$ & 327.8 $\pm$ 63.4$^\dag$ & 199.4 $\pm$ 14.2$^\dag$ & 525.8 $\pm$ 11.7 
\\\bottomrule
\end{tabular} }
\label{tab:res:tilt}
\end{table}}

\section{Discussion} \label{Sec:discussion}
\noindent
This study proposes a method for estimating AV node conduction properties with beat-to-beat resolution during AF utilizing a network model of the AV node and a particle filter together with a smoothing algorithm. The method was evaluated using simulated data to analyze how well the particle filters estimated the AV node conduction properties for each heartbeat. Additionally, synchronized EGM and ECG recordings were analyzed to study the concordance between estimates obtained from the two data types. Finally, ECG recordings obtained during a tilt test protocol were analyzed to evaluate the method's ability to quantify expected changes in AV node characteristics. \newline
\indent
The proposed framework builds on a physiological model where $\boldsymbol{\phi}$ must be estimated for each heartbeat based on noisy observations. In this setting, optimization-based techniques such as a genetic algorithms could be used to estimate $\boldsymbol{\phi}$ beat by beat, but these would only yield point estimates and not the posterior. While Markov chain Monte Carlo methods, such as Metropolis-Hasting, could approximate the full posterior, they are too slow for dynamic beat-to-beat inference. Alternatively, recursive Bayesian methods such as the Kalman filter are computationally efficient but rely on strong assumptions, such as linear dynamics and Gaussian noise, which are not applicable here. Particle filtering and smoothing, by contrast, allow for efficient approximation of the full posterior in a nonlinear, non-Gaussian, and time-varying system -- making them well-suited for this application.\newline
\indent
The estimation accuracy of the present study can be compared to results from our previous study, where the same AV node model was used to estimate $\phi$ with a 10-minute resolution using an approximate Bayesian computation approach \cite{karlsson2024model}. In that study, $\phi$ could be estimated with a mean absolute error of 111 ms for $R^{FP}$, 12 ms for $R^{SP}$, 90 ms for $D^{FP}$, and 110 ms for $D^{SP}$. The present study, by contrast, introduces beat-to-beat tracking of AV nodal dynamics, which represents a major advancement in temporal resolution and allows for the analysis of short-term variations that were not accessible in the earlier framework. \newline
\indent
Comparing these results to the ECG $\overline{l^1}$ and ECG $\overline{l^1_{min}}$ in Table \ref{tab:res:GT}, it is evident that the beat-to-beat estimates in the present study have a higher error (ECG $\overline{l^1}$) for $R^{FP}$, $D^{FP}$, and $D^{SP}$, and a drastically higher error for $R^{SP}$, illustrating a trade-off between temporal resolution and estimation accuracy. Conversely, the error of the one-minute averaged estimates (ECG $\overline{l^1_{min}}$) is comparable to the error in \cite{karlsson2024model}. The computation time in the present study (17.5 minutes per minute of ECG) was multiple magnitudes greater than the 40 seconds per minute required using the framework of the previous study. Hence, the proposed methodology enables a more detailed analysis of the AV node dynamics at the expense of increased computational cost. \newline
\indent
The accuracy of the ECG-based estimates differs between AV node conduction properties, with the lowest error obtained for $R^{SP}$ (6\%) and the highest for $D^{SP}$ (16\%), as seen in Table \ref{tab:res:GT} (ECG $\overline{l^1}/\boldsymbol{r}$). Comparing the estimated refractory period trends to clinical data is difficult, since the AV node refractory period is assessed using pacing protocols resulting in a point estimate rather than a trend. In addition, pacing protocols assess the refractory period under controlled conditions, not during the irregular conditions of AF, as is the case in this study. Nevertheless, the results in Table \ref{tab:res:GT} suggest that the model and framework cannot identify variations in the $R^{FP}$ and $R^{SP}$ smaller than 169 ms and 67 ms, respectively. \newline
\indent
The differences in estimation accuracy across AV node properties arise from both signal characteristics and model dynamics. Fast pathway properties are estimated with higher error for the simulated data since fewer beats conduct through it, reducing its affect on the output as well as the data available for inference. Moreover, $R^{SP}$ is estimated with similar accuracy using ECG-PF and the EGM-PF, as seen in Table \ref{tab:res:GT}, likely due to its strong influence the timing of RR intervals and can therefore be inferred reliably even from ECG recordings. In contrast, the conduction delay is more accurately estimated from EGM recordings, since they provide direct information on atrial activation that is not visible in ECGs. \newline
\indent
Only a few sources exist for functional and effective refractory periods in the SP and FP of the AV node in humans. In the study by Denes et al., the functional refractory period for two patients with paroxysmal supraventricular tachycardia and evidence of dual AV nodal conduction was 820 ms and 495 ms in the FP and 540 ms and 414 ms in the SP \cite{denes1973demonstration}. In the study by Blanck et al., the effective refractory period in 18 patients with inducible sustained atrioventricular nodal reentrant tachycardia varied from 230 ms to 440 ms in the FP, and from 180 ms to 420 ms in the SP \cite{blanck1995characterization}. Assuming that intra-individual variations in the AV node refractory period are smaller than the difference between individuals, comparing the reported results to the results in Table \ref{tab:res:GT} indicates that identifying beat-to-beat variations in $R^{FP}$ is generally not possible. Instead, the one-minute average, with an error of 92 ms, may be sufficient to detect transient changes in $R^{FP}$. Moreover, the beat-to-beat error of 67 ms suggests that the model and framework may be sufficient to identify beat-to-beat variations in the $R^{SP}$. Furthermore, the results in Table \ref{tab:res:GT} suggest that variations in the $D^{FP}$ and $D^{SP}$ smaller than 178 ms cannot be identified. Beat-to-beat variations in AV conduction delay of such large magnitude are generally not observed during normal sinus rhythm \cite{denes1973demonstration}. However, it cannot be excluded that larger variations in conduction delay may occur during AF. \newline
\indent
\begin{figure*}[b]
    \centering
        \includegraphics[width=18cm]{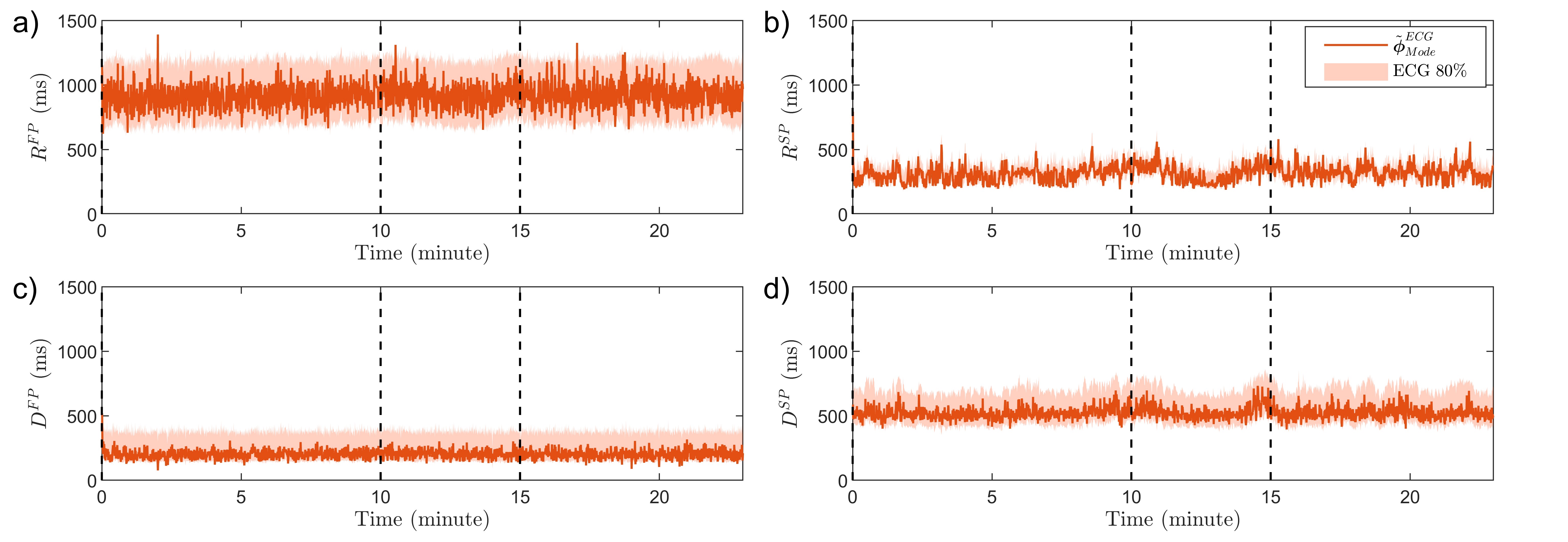}
    \caption{The estimated modes $\tilde{\boldsymbol{\phi}}^{ECG}_{Mode}(k)$ (lines) and the 80\% credibility region (shaded background) for one patient in the tilt test study. Panel (a) show $R^{FP}$, (b) $R^{SP}$, (c) $D^{FP}$, and (d) $D^{SP}$. Dashed vertical lines indicate the time of tilting, starting in supine position, following five minues in HDT, before HUT. }
    \label{FigureY}
\end{figure*}
The iafdb was used to study the concordance between estimates obtained from synchronized ECG and EGM recordings. As seen in Figure \ref{Figure4}, there are some differences when using EGM recordings compared to ECG recordings. This is clearest in $D^{SP}$, where estimates obtained using EGM recordings have higher beat-to-beat variability than those obtained using ECG recordings. Thus, the ability to track fast changes is somewhat reduced when only using non-invasive recordings. A difference in convergence time also exists, most clearly visible for $R^{FP}$ in Figure \ref{Figure4}. Such convergence behavior was seen for all patients in the iafdb database. Nevertheless, the estimates for all four AV node properties are similar using PF-EGM and PF-ECG, with no average difference larger than 5\%. \newline
\indent
The AV node conduction properties estimated based on the ECG recordings from the tilt test study, summarized in Table \ref{tab:res:tilt}, show a significant decrease from supine to HUT position for $R^{FP}$ ($p < 0.05$), $R^{SP}$ ($p < 0.001$), and $D^{FP}$ ($p < 0.01$). Changing the position from supine to HUT is associated with increased sympathetic activity \cite{aponte2021tilt}, and hence a decrease in AV nodal conduction delay and refractory period, as previously discussed in \cite{plappert2022atrioventricular}. Thus, the results summarized in Table \ref{tab:res:tilt} are in line with what can be expected for HUT, suggesting that the proposed methodology enables the detection of tilt-induced changes in AV node conduction properties. Although the decrease is significant and in line with what is expected for the studied population, the differences are smaller than the corresponding one-minute averaged error (ECG $\overline{l^1_{min}}$) seen in Table \ref{tab:res:GT}, limiting the possibility to detect the changes on an individual level. In addition, the resulting RR interval is influenced not only by the AV node properties, but also by the electrical activity in the atria \cite{mase2015dynamics, climent2010role}. For the tilt test dataset specifically, a significant increase compared to baseline in the atrial fibrillatory rate was observed during HDT \cite{ostenson2017autonomic}. Hence, variations in the RR interval across tilt phases cannot be attributed solely to changes in AV node properties, complicating the analysis. \newline
\indent
Moreover, it is less clear how changing the position from supine to HDT affects the sympathetic and parasympathetic activity, and the results from our analysis do not show any significant changes in AV node conduction properties between supine and HDT. Further, as suggested by the estimated trends of $R^{SP}$ and $D^{SP}$ in Figure \ref{FigureY}, the proposed methodology may enable the detection of transient changes in the AV node conduction properties. \newline
\indent

\subsection{Study limitations and future perspectives}
\noindent The network model accounts for several important properties of the AV node conduction during AF, however, it is by no means a perfect replica. For example, it does not include ventricular escape rhythm and the network topology used in this work excludes some uncommon AV node structures such as multiple slow pathways. Nevertheless, these simplifications are essential to developing a model with a manageable number of parameters. Moreover, seeing as these are uncommon structures, these limitations are not likely to affect the results. \newline
\indent
The estimated AV node properties have only been validated using ground truth data generated from the same AV node model. However, obtaining the exact refractory period and conduction delay in both pathways from patients suffering from AF -- if at all possible -- would be very difficult and time-consuming. In addition, the number of patient recordings analyzed in this study is limited, which restricts the ability to generalize the findings across the wider AF population. \newline
\indent
While a computation time of 17.5 minutes per minute of analyzed data limits the clinical application of this framework, this could be addressed in several ways. Since each particle in the particle filter is computed separately, making the method parallelizable across CPU cores would reduce processing time. In addition, the number of particles is directly proportional to computation time. In this study, 1,000,000 particles were used to explore the potential of the framework to study beat-to-beat variations. Many applications, such as comparing tilt positions in the tilt test data, do not require beat-to-beat resolution. In those cases, using fewer particles and basing the analysis on one-minute averages could be sufficient. However, to study the transition period between tilt positions, beat-to-beat resolution is necessary. The number of particles, and thereby the computation time and accuracy, needs to be chosen to match the requirements of the clinical application.\newline
\indent
Moreover, since inter-patient variability in ANS activity might influence individual responses to AF treatment, it would be interesting to quantify the ANS activity with the proposed framework during common AF treatments such as different rate control drugs.

\section{Conclusion} \label{Sec:conclusion}
\noindent 
We have proposed a novel framework for estimating patient-specific AV node properties with beat-to-beat resolution and conservative uncertainty estimates utilizing a mathematical model combined with a particle filter and smoothing algorithm. By using synchronized EGM and ECG recordings for the parameter estimation, the loss of estimation accuracy using non-invasive recordings could be studied, suggesting a sufficient accuracy for capturing beat-to-beat changes in the refractory period of the SP. \newline
\indent We illustrate the potential of the proposed methodology by analyzing a tilt-test dataset. Results suggest that changes in AV node conduction properties can be assessed from ECG using the proposed method.

\section*{Conflict of Interest Statement}
The authors declare that the research was conducted in the absence of any commercial or financial relationships that could be construed as a potential conflict of interest.

\section{Author Contributions}
\noindent MK, FS, and MW contributed to the design and conception of the study. SÖ performed the clinical study generating the tilt test data. FP extracted the RR series and f-waves from the ECG for the tilt test data. MK wrote the manuscript, particle filter, and smoothing algorithm, with advice, suggestions, and supervision from FS and MW. PP and SÖ analyzed and interpreted the results from a medical viewpoint. FS and MW supervised the project and reviewed the manuscript during the writing process. All authors contributed to the manuscript revision, read, and approved the submitted version.

\section*{Acknowledgment}
The computations were enabled by resources provided by the National Academic Infrastructure for Supercomputing in Sweden (NAISS) at Tetralith partially funded by the Swedish Research Council through grant agreement no. 2022-06725.

\section*{Funding}
\noindent This work was supported by the Swedish Foundation for Strategic Research (Grant FID18-0023), the Swedish Research Council (Grant VR2019-04272), and the Crafoord Foundation (Grant 20200605).


\section{Data Availability Statement} \label{Sec:DAS}
\noindent The estimated AV node properties and the simulated data supporting the conclusions for this article will be available from MK upon request. The code for the model together with a user example can be found at \url{https://github.com/FraunhoferChalmersCentre/AV-node-model}. The iafdb data are publicly available on \url{https://physionet.org/content/iafdb/1.0.0/}. The tilt test protocol data is owned by the Department of Cardiology, Clinical Sciences, Lund University, Sweden, and access requests should be directed to pyotr.platonov@med.lu.se.

\bibliographystyle{IEEEtranN}
\bibliography{Ref}

\end{document}